**Towards Non-Linear Cultural Production and Systems of Machinic Agency:**
**In the Case of TikTok Value Generation**

H. "Kevin" Jin

kevin.jin@student.uva.nl

# Abstract

The rise of TikTok has brought forth novel ways to create and consume media content, accelerated by technologies such as hyper-individualised algorithms and easy-to-use video production tools. Despite its popularity, scholars and politicians alike have raised many concerns on the legitimacy and ethics of TikTok regarding its services, and its collected data. However, much of these discussions take the premise of user-generated content for granted, attributing them to human expression without critically evaluating how the *making* of on-platform content production have changed. With a grounded theory approach, in conjunction with a platform-aware walkthrough that pays special attention to *the material and immaterial premises of platform value generation,* my findings suggest that the intensification of datafication have proliferated from consumption behaviours to the process of content production, whereas content production no longer solely produce media content. As platforms become the active recruiter, mobiliser and co-producer of media production, I argue that it is no longer feasible to distinguish human and machine contribution in the ways they are consumed to facilitate platform valorisation. I propose that the technical arrangements of TikTok, in relation to its users has fostered a *non-linear* mode of platform cultural production capable of generating economic value through a system of *machinic agency* that incorporates human and machines in an indistinguishable manner. As content, the premises of platform valorisation has become an inseparable effort of human-machines, I urge that the relationship between technology and humans be reassessed as a system of machinic agency that mutually shapes our mediated reality, rather than singular, differentiable actors that contribute to platforms.

# Table of Contents

# Introduction

We are no strangers to very large tech companies, their "platforms", and their so-called "free" services of sorts. YouTube videos appears to be free to watch and upload; X (previously Twitter) seemingly affords users to create and interact with others through threads free of charge; Google makes its search engine openly available. A more recent player amongst these "tech giants" is TikTok: ever since its worldwide launch in 2018, their software has amassed over 2 billion downloads, gaining significant user bases across the globe. As a social media "platform" embedded with technologies such as hyper-individualised algorithms and easy-to-use video production tools, TikTok not only affords expression to millions of users through short-form videos, but also serves as a centrum for personalised media consumption for many more. TikTok has become an emergent venue for individuals to experience and engage with the world through its technological architectures and contents—manifesting a reality where information is by and large mediated through many actors beyond themselves; a world that permeates from one screen to another, regardless their location or time. Users have quickly learned to navigate and mobilise TikTok's services, as we soon find TikTok permeating through public spheres, anywhere from presidential elections to personal relationships (cf. Paul 2022; Şot 2022). TikTok is distinct in its ways of content circulation: apart from watching and sharing videos on TikTok, the application allows users to directly shoot, edit and distribute video content—all without the need to open other programme. Edits and effects that previously take much learning to achieve in specialised softwares[1] are now accessible on TikTok with merely a few touches and swipes and no prior knowledge. The power of media production previously exclusive to professionals and hobbyists seemed to have proliferated to a much larger public on TikTok thanks to its integrated system. As TikTok integrates the content production process in its application, it gains even more significance in our communication and cultural circulation by becoming a self-fulfilling cultural environment that creates, distributes and circulates media content all within itself (Zhang 2021). But why is TikTok designed in such a way? As platforms host, connect and converge individuals, institutions and *other platforms* alike, how are we situated in their operations? As we use TikTok and many other platform services through their applications, how do their profit-driven nature affect us in our *mediated reality*? As Poell,

---

[1] such as Adobe After Effects and Apple's Final Cut Pro.

Nieborg, and van Dijck (2019) highlighted, platforms warrants careful consideration not only regarding their cultural impacts as it may reorganise "cultural practices and imaginations around platforms", but also their political economy footprints as intrinsically asymmetric power relations between "operators, end-users, and complementors" configured by dominant corporations and amplified with network effects (5-6). Their observation is particularly important when it comes to social media platforms that profits from circulating discursive content and actively contributes to what individuals experience as culture.

Driven by these curiosities, this study empirically traces the relation between users and the platform through TikTok's software interface during the process of content-making. I highlight this particular process as it hones twofold significance for both TikTok as a *business* and its *users*. For a corporation that circulates user-generated content, media production dictates what can be consumed, which in turn allows for other profit-making operations such as advertising and commission (Mhalla, Yun, and Nasiri 2020; Ma and Hu 2021). For users, creating content on TikTok is a social and expressive practice that connects themselves to others: their contents have consequences beyond themselves, as they become what other users perceive and experience on the platform (Collie and Wilson-Barnao 2020). With a keen focus on this particular exchange between users and the platform, this thesis is guided by an overarching research question: *How does TikTok configure and transform the process of content production to generate value for the platform?*

To answer this question, I conduct a data-driven, platform-aware walkthrough on TikTok's content production interfaces, supplemented by investigations on the platform's documentations and access points for other stakeholders, such as advertisers and creative professionals. The processes of content production are interpreted as both cultural practices and technological interactions in the context of platformisation. This framing allows me to be analyse TikTok's configuration both from a focused perspective that affects individuals' communication practices, while addressing the greater cultural and economic consequences of such arrangements amplified and intensified with the networked nature of TikTok as a profit-driven platform. I will demonstrate that by designing an accessible yet functionally constrained interface that empowers data capture and circulation, TikTok is able to generate value by capturing users and subsuming their contribution in the form of data. I suggest that content itself hones no value, but TikTok has configured content and its surround practices to *extract* value from users. Conceptualising my findings through a grounded theory approach, I suggest that valorisation through *datafication* on media platforms have proliferated from *consumption* patterns and content aggregation to the practices and knowledge emerging during the process of content *production*. I will argue that, through intricate data circulation, TikTok has fostered a *non-linear* mode of cultural production where the discursive media is no longer *all that there is* of content, nor is it the only artefact circulated to pose cultural influence. I will also argue that content produced on social media platforms has become increasingly an indistinguishable collaboration of human and machines produced through a system of *machinic agency* rather than actors in isolation, which raises critical concerns on the much-favoured Marxist theory of value that attributes human labour as the sole source of value. I conclude that through the convergence and datafication of previously discrete processes of cultural production for the purpose of platform value generation, the notion of algorithmic culture has proliferated from cultural consumption to production practices with TikTok's algorithmic and automation centred production interfaces with ultra-short feedback loops geared towards corporate interests, resulting in consequences that are yet to be known.

In the following sections of this thesis, I outline how content production and its consequences should be scrutinised in the context of platformisation and value generation. Then, I explore on existing literatures and offer a conceptual inventory that address production as both a cultural practice and an interaction between the user-platform that could be configured through the user interface.

## Theoretical Approaches

To tackle our research question, we need to first understand how platforms generate value in relation to its users, architecture and other connected parties. We will pay specific attention to the material and immaterial premises that facilitate platform valorisation. We then describe how content is intricately tied to these premises and contribute to platforms' value-deriving potential, specifically through user retention and data generation. We outline that the contribution of content lies in its power to extract value from users under platform circulation mechanisms, rather than its explicit potential to generate monetary value.

*From Platformisation to Platform Valorisation*

By referring to TikTok and other social media providers as *platforms*, scholars and the public acknowledge that they encompass more than their utilitarian affordance of sociality and communication, and connect to various parties other than end-users. TikTok, for example, is at the same time a stage of user-generated content, an advertising agency, a matchmaking service for business and creators, and a distribution channel for professional creators. With its aggregation of functionalities comes its extensive ability to condition, configure and control a multitude of activities of its intersecting parties (Hein et al. 2020; van Dijck and Poell 2013). Behind this convergence, we must keep in mind that TikTok is ultimately a profit-driven private company that have to answer for its shareholders; its operations are thus ultimately orchestrated towards corporate interests. As many scholars have promptly noticed, most platforms and their services are far from neutral, nor as "free" as they seem due to their elitist ownership status and revenue-oriented nature (cf. Lutz et al. 2021; Duffy and Wissinger 2017; Nieva 2015; Gillespie 2010). But how do social media platforms generate value? Scholars across the social sciences and humanities have largely agreed that most corporate-owned media platforms profit from the collection, storage and processing of user information and commodifying them either in its data form or as a service, such as advertising (cf. Akmeraner 2018; Duffy and Wissinger 2017; Plantin and Punathambekar 2019). For the information-rich social media, valuable information about users' interests, preferences, social and economic status can be captured following users' engagement with content, which could be used to optimise platform service as well as generating surplus value. TikTok affirms in its ToS and privacy policy that user information from various sources are processed and shared for advertising and other monetary purposes ('Terms of Service | TikTok', n.d.). A close reading on TikTok's ToS also reveals other possibilities of platform valorisation that differentiates its valorisation model from other service providers. Apart from offering the deployment of advertisements, sponsorships and campaigns as a *service*, where marketers and advertisers pay the platform in exchange for exposure and engagement, TikTok also earns commissions, particularly via users' purchase of virtual gifts to content creators, as well as e-commerce activities when users buy products from the platform's shopping feature. Despite the variety in valorisation methods, they all to a great extent rely on the circulation of content: advertisements primarily takes the form of "pre-rolls" displayed before or in-between content feeds; sponsored content are circulated next to normal user-generated videos; whereas in-app purchases and e-commerce features relies are mostly driven and leveraged by creators' content appeal to users either as livestreams or short-form videos (Shi 2019, Li and Lo 2015, Campbell et al. 2017). Amongst them, targeted advertisement has prevailed in its utilisation on TikTok (Chen 2017). From here, it could be concluded that TikTok's value generation is largely dependent on the active creation, circulation and consumption of content.

*Does Content Have Value?*

Although content hones substantial significance, we must bear in mind that content does not emerge from thin air, nor can it generate value by itself (as content is offered for free consumption); it only yields value in relation to the specific arrangements and logics of the platform. In order to understand how content relates to these arrangements and how their production brings platform value, we turn to political economy critiques that highlights the material and immaterial premises of valorisation following their research on the subjects and mechanisms of exploitation.

Poell, Nieborg, and van Dijck (2019) identified three structural dimensions of platforms essential to sustaining their operation; namely, *data infrastructures* that captures and transforms information into data ("datafication"); *markets*, and *governance*. These dimensions, in turn, allows platforms to capture and transform user activities into value.

Returning to targeted advertising, the dominant source of revenue on TikTok and many other platforms, we need to keep in mind that value generation through this method is achieved through mediated connections of multiple parties: there must be an advertiser paying for the service, users that create and consume content, and platform hardwares facilitating content consumption and advertisement deployment. As such, TikTok's valorisation depends its ability to appeal to these stakeholders, specifically through the following mechanisms:

- **Attract and retain users through content consumption and generation.** By keeping as many users as possible and hosting them as long as possible, the platform occupies and harvests users cognitive functions, thereby turning users' attention into usage data and subjects of paid promotion exposures (Celis Bueno 2017). Alternatively, by creating content, user develop interpersonal connections on the platform and provides more media content to be circulated that develops sociality between users and encourages consumption. An extended user demographics with prolonged usage signifies *user adhesiveness* to the platform, which increase competitiveness of the platform's advertising service (Li, Li, and Cai 2021). For social media platforms, attracting users and their usage is closely related to the number of existing users and their connections established on the platform, whereas connections are manifest through content (Aamir et al. 2024; Bucher and Helmond 2018). Generally, the more users and connections there are, the more likely it is for the platform to attract more upcoming users and content (Currier 2019). This relation has been identified as *network effects* that allows the value or utility of platforms to grow exponentially and gravitate to their service (Van Der Vlist 2022, 27).
- **Extract data from users.** Both TikTok's content feed function and its targeted advertising require the platform to extract user-specific information, commonly by deriving relations between users' consumption patterns and content. Such extraction is warranted under platforms' legally binding terms-of-service (ToS), and technologically enacted by simultaneously processing media content and its semantics to data, while capturing users' consumption behaviours. This process reflects a broader tendency of *datafication*, through which platforms capture and "render into data many aspects of the world" to facilitate their operations (van Dijck, Poell, and Waal 2018, 33).

It should be noted that these two aspects functions in synergy: more users on the platform gives the possibility of generating more connections through content, whereas with more content allows for more data to be generated during consumption, which in turn enhances the platform's features to curate individualised content feeds to attract and retain usage and deploy advertisements. We can find that content is neither the beginning nor the end for platform valorisation, but rather one step instrumentalised within the interwoven operations of TikTok to generate value. *The value of content, on TikTok and many other social media platforms, therefore lies in its utility to extract value from users, whether that is creative labour, attention, or their overall experiences and intellect* (cf. Fuchs and Sevignani 2013, Collie and Wilson-Barnao 2020, Zuboff 2020, Gandini 2021).

These insights allow us to interpret *what* TikTok attempts to achieve by configuring content production, namely user retention and data generation; regarding *how* the practice of making content can be configured by the platform, we turn to a curated walkthrough approach to explore how users actions are arranged and constructed through a specific sociomaterial environment, that is, TikTok's` software interface that affords users' content production.

## Methods

*A Data-Centric, Platform-Aware Walkthrough*

Informed by research pragmatics in platform and software studies, this investigation highlights the software user interface, as an alternative entry point to understand platform mechanisms. I propose a *data-centric, purpose-driven walkthrough* method to systematically examine how the TikTok mobile application configures users' content production practices, and trace the ways in which the TikTok converts, translates and presents data captured during users' interactions with the software. This method takes into account the software's multifaceted nature that expects various usages that features functionalities beyond content production, and attends on particular usages instead of others to map software operations with platform mechanisms on a granular level.

The interface hones twofold importance for our study: it is both a intentionally designed system provided by TikTok that consequently embeds and operationalises its corporate interests, and a manifestation of actions, networks and their consequences between the user, their peers and the platform architecture, as "forms of relations" (Hookway 2014, cited in Bucher and Helmond 2018). To study the interface is thus to study the intention of TikTok through its interactive designs, and how such a technological arrangement interplay with user activities through the native language of digital technologies: data. In practice, our method is primarily inspired by the walkthrough method outlined by Light, Burgess, and Duguay (2018), designed to analyse software applications and user-software interactions. Building from Latour's Actor-Network Theory (2005), affordance theories (Bucher and Helmond 2018; Norman 1999; Gibson 1979) as well as infrastructure and culture studies, Light et al. acknowledges software applications as technological actors posses agency, as they could affect users and guide their activities with designed interfaces as environments for actions, whereas the arrangements within technological systems form a "master narrative" that could encode presumptions and cultural aspirations from their creators. Light's original walkthrough method was designed as a toolkit to comprehensively examine software applications by the context from which they were designed and created, such as their "vision, operating model, and governance" (10), combined with their empirical workings, specifically how the application interface "transform meaning through the interactions they invoke" (14). Through a step-by-step recording and observation of softwares "screens, features, and flows of activity" during "registration and entry; everyday use; and suspension, closure and leaving" (3, 14), researchers can analyse the technological arrangements of applications as cultural discourses that configures users actions. However, in the case of TikTok as well as many other softwares, the interface is no longer a static artefact that responds to user inputs without variations: the interface, as the visible, interactive part of a greater technological system, is now capable of producing unpredictable outputs as the technical architecture behind it evolves and accumulates data. While the unpredictability of the interface (or, what information the interface represents) has been regarded as a challenge in some interface studies (Duguay and Gold-Apel 2023), it is precisely this differentiated processing of inputs that demonstrates how actors of a platform could "also afford things back to the technology" through networked mediation (Burgess et al. 2016, 4). While we acknowledge Light's identification on the "environment of expected use" (3-4) before beginning a technical walkthrough, it could be possible that the application expects multiple usages that each requires their own exploratory approach. In the case of TikTok, consuming and creating content as well as direct messages are all expected uses of the application, yet we can quickly discern that the interface presented for these usages are fundamentally different. Additionally, although Light's methodology yields comprehensive empirical data through its step-by-step observation, exhaustively exploring every stage of "registration and entry, everyday use, and discontinuation" of an application's expected uses results in a large amount of forensic evidence that might risk a dispersion of research focus and challenge conclusive contributions (ibid., abstract). When applied to TikTok, the original method generates data that is somewhat strictly end-user oriented, as the software primarily serves consumers. In 2023, Duguay and Gold-Apel evaluated the usefulness of walkthroughs on emerging platforms such as TikTok, noting their "increasingly complex platform ecology warrant consideration of supplemental and alternative methods." (8) Indeed, TikTok's operation is not merely limited to end-users (B2C) but also business partners (B2B). As such, simply inquiring in user interactions without considering its wider implications in a multisided platform could not sufficiently account for the role of TikTok in our current socio-economical setting. Given the alert from Duguay and Gold-Apel (2023), the data yielded from my walkthrough is supplemented by explorations and close reading of TikTok's business-oriented resources, including the "TikTok for business" website and its advertising dashboard, "Ads Manager" to further understand the ways TikTok configure and incorporate user activities in its profit-yielding ecosystem.

Considering my interest in content production on TikTok in relation to the platform's mechanisms of valorisation, I reorient the walkthrough such that I trace user interactions given the specific "expected use" of content production to empirically capture how user activities are configured on the platform to for the purpose of data generation and user retention. I refer to *content production practices* as user activities that contribute to the creation of audiovisual content that could be circulated on the platform as videos. In such case, we do not extensively analyse or discuss the practices involved when using TikTok to create other forms of media content such as Stickers or Effects, nor other possible usages of TikTok, such as direct messaging.

Based on preliminary explorations on TikTok's content production process, my walkthrough adapts Light's distinction between various stages of application usage to the specifics of media production, resulting in three stages of production on TikTok: *entry*, *production* and *distribution*. The **entry** stage focuses on what *leads to* making content: that is, how interactive elements and their underlying data are employed to initiate and provoke users towards content production. **Production** focuses on the stage where users activities (including actions and media inputs) are transformed into TikTok content. We pay specific attention to the ways in which user actions might be influenced by actors other than themselves. Finally, the **distribution** stage marks the exit of content production by users contributing finished content to TikTok in the form of both media and data. I attend to the data that TikTok attempts to capture apart from the audiovisual media itself, and the interface narrative that facilitates data capture and transformation. To substantiate my interpretation of the interface, I refer to support manuals on TikTok's website, where the usages of different tools and features were explained. These documents not only reduces ambiguity that might occur during our interface exploration: their rhetoric regarding the software are also usefully symptomatic of the company's imaginaries and expectations behind the interface. Additionally, the utility of content production configurations is evaluated in accordance to empirical investigations on the platform's documentations and access points for other stakeholders, such as TikTok's Business dashboard[2], Creator Academy[3], Effect House[4] and CapCut[5], respectively intended for business partners and creative professionals.

*Challenges: Ethics and Data Preservation*

Earlier TikTok studies have warned us that the software and its interface might undergo rapid changes with little prior notice, as Duguay and Gold-Apel (2023) referred to this phenomenon as "Ephemeral Interface Instantiations" (7), and noted such changes might challenge researchers to accurately capture data from the software, thereby undermining their research contribution. Keeping in mind this inherent ephemerality of digital artefacts surrounding TikTok, this project records all primary evidence for digital artefacts from TikTok that were referred in this study, utilising illustrations, screenshots, recordings and web archives to effectively preserve the context of our discussion in case of future changes to the platform. Since TikTok circulates content generated by users, it is often unavoidable for researchers to encounter pages within the software that contain sensitive information of other creators. Nevertheless, depictions of those screens could still be an essential supplement to our analysis of the TikTok interface. Simply modifying the collected screenshots to censor data is not enough-as we risk redacting interface elements essential to our investigation. As an effort to visually communicate our analysis with respects to user privacy, mockups illustrating the interface are created instead of using screenshots to depict the software interface for pages that potentially contain users' personal information.

---

[2] available at https://www.tiktok.com/business/

[3] available at https://www.tiktok.com/creator-academy

[4] available at https://effecthouse.tiktok.com/

[5] available at https://www.capcut.com/

*Grounded Theory Approach*

My findings following the software walkthrough is interpreted using a *grounded theory approach* that attempts to derive and formulate theories from empirical exploration, comparison and abstraction of empirical interactions (Glaser and Strauss 2017). I employ this inductive approach to identify emerging concepts and mechanisms that may arise from my exploration of the user interface. Inspired by Zulli and Zulli's (2022) pioneering study on TikTok's memetic properties using a similar approach to conceptualise a "imitation publics" building on notions of networked publics, I intend to demonstrate that novel phenomena have arisen from the platform's content production configurations that allows for theories of cultural production and agency to be extended.

# FINDINGS

This section reports on the technological and design arrangements of TikTok's software interface that facilitates content production through various stages.

*Entry*

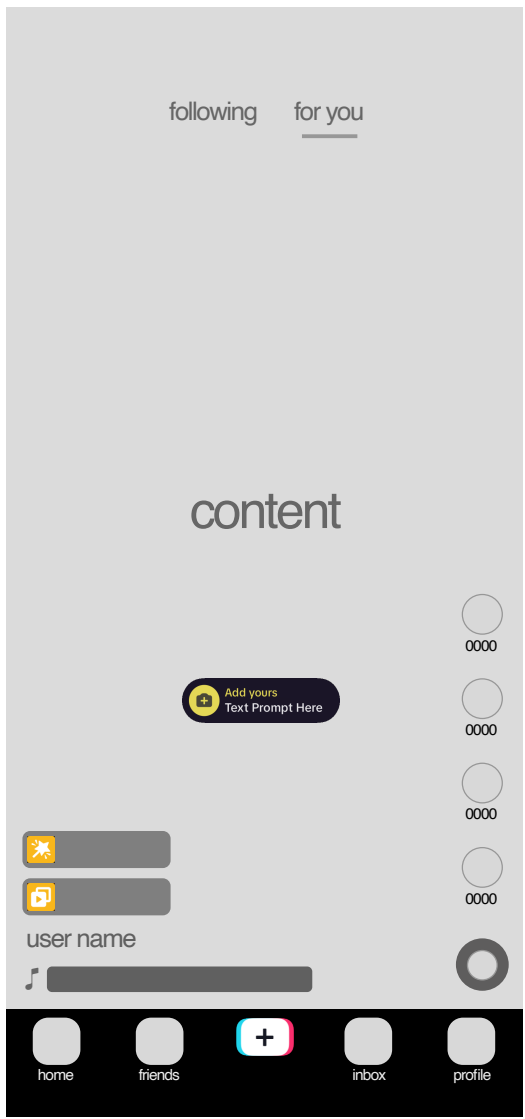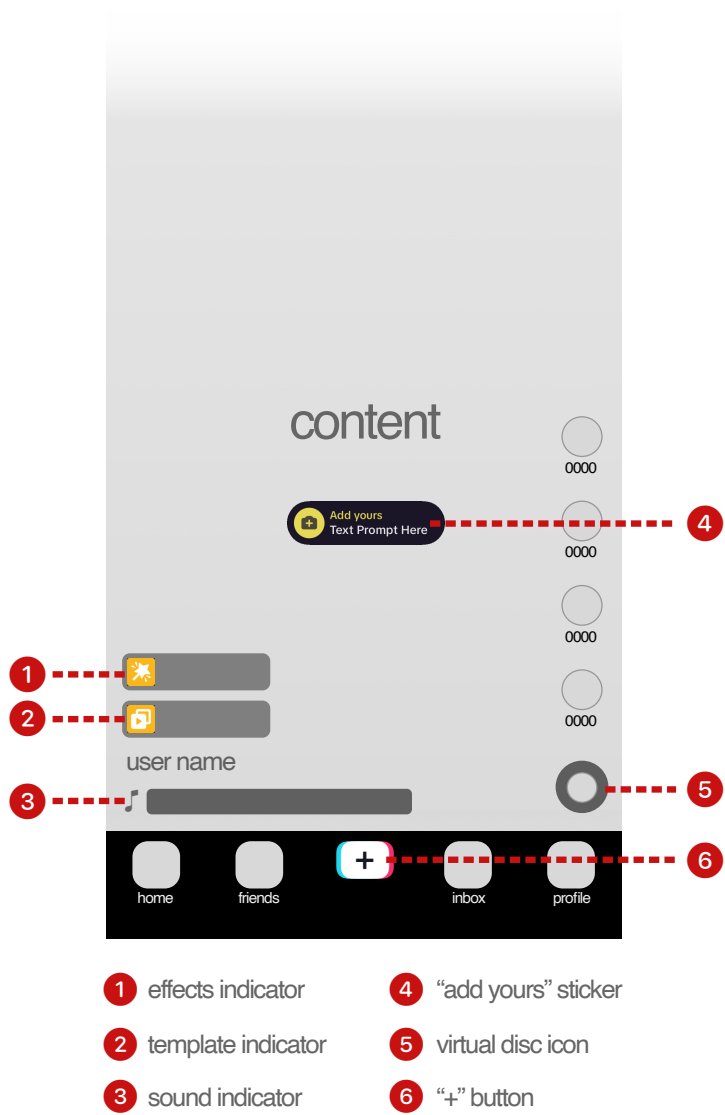

**Figure 1**. Annotated mockup of the TikTok Home screen

On TikTok, there are several entry points for users to initiate the content production process. **Figure 1** depicts and annotates the home page of TikTok that greets users upon launching the application. Apart from pressing the "+" button located on the centre-bottom of the home screen, users can also enter the production process while consuming content either by pressing the *virtual disc icon* on the bottom-right corner overlaying on the content to produce content using the same sound as their currently playing video, or pressing the *effects indicator* overlay on the bottom-left, if the content uses existing effects on TikTok. When users press the *virtual disc icon*, or the *effects indicator*, they are brought to a screen that shows details about the sound or effects, including its author, popularity, related content using that sound or effect, in addition to a button that prompts users to "use this effect/sound". From the sound/effects detail page, users can press on the prompt to enter production. Alternatively, the user can enter production through the *template indicator* overlay on the bottom-left when consuming content using TikTok templates. In this case, users would enter the template preview screen that prompts them to input media content. With a closer look at the signifiers employed at these entry points, we see that TikTok renders content production as a participatory activity for its users: the "Add Yours" prompt explicitly encourages user participation, whereas effects, template and sound indicators function as shortcuts to content production using existing audiovisual and stylistic resources. By situating multiple production entries as indicators on the home screen of TikTok, producing content becomes a noteworthy action possibility for users that is repeatedly signified on the interface.

*Production*

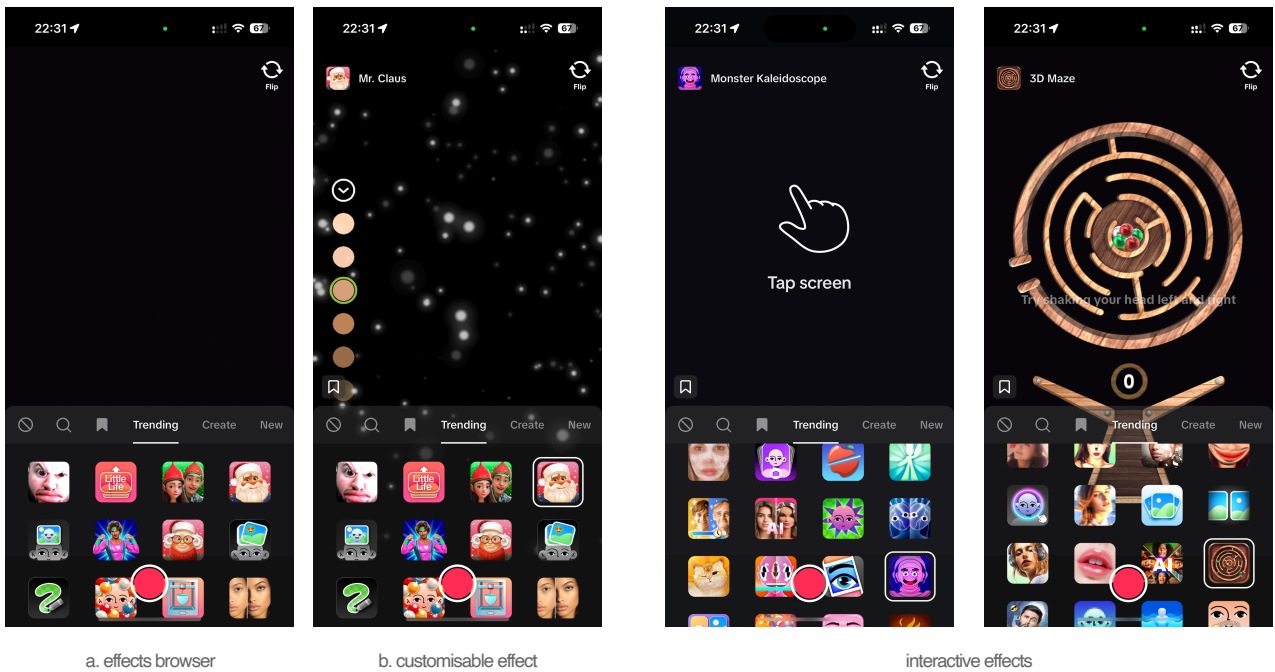

**Figure 2a and b.** effects selector  **Figure 3**. Interactive effect examples

After *entry* comes the stage of *production*, where users gather audio and visual media resources either by recording live or uploading existing media from device storage. On this page, a set of recording controls is accessible to the user on the right side of the screen, including camera controls, timer, filters and "retouch" that primarily modifies facial features during live capture. The recording button at the bottom-centre of the page toggles between start and pause of the camera recording. If users did not previously select an effect during the entry stage through the sound/effects detail screen, users can browse a range of effects that modifies their recording by pressing the "effects" icon on the bottom-left (**Figure 2a**). Created by TikTok and other users, these effects can change the camera feed to various degrees: some modifies facial or body features, some are customisable (**Figure 2b**), some provides interactivity (**Figure 3**) and responds to users behaviours, whereas others attach sounds to the recording. In consultation with TikTok's business and creator oriented websites, it should be noted that resources options such as sound, effects and templates are not always made available by TikTok, but rather an aggregation from three primary sources: end-users, creative professionals, as well as business sponsors (TikTok 2019; n.d.): end-users may choose to publish their content as a template (as I will further elaborate in the distribution stage), whereas creative professionals can create templates and effects respectively using CapCut and Effect

House, two desktop-based professional production softwares offered by TikTok. Businesses can also deploy sounds and effect assets as "Branded Effects" to encourage brand engagement and exposure during the process of content production (TikTok, n.d.; Mhalla, Yun, and Nasiri 2020; Ma and Hu 2021). Users can also record with multiple effects by pausing their capture and selecting another effect. There seem to be virtually no limits on how many effects users could include in their recording[6]. Since these effects are only applicable during live recording, I will refer to them as *Live Effects* hereafter to avoid ambiguity with the Effects available in the following pages. By tapping on the "Add Sound" button on top of the screen, users enter the *sound selection page,* where they can choose sounds from TikTok's sound library, or extract sounds from on-device audio or videos. Regardless the audio source, only one piece of sound could be added on this page in addition to the live-recorded audio. The selected sound is only applied to live captures and played while recording is active, which allows users to sync their live actions with a soundtrack. When users press "upload" located at the bottom-right of the page, they can consequently select and import existing photos and videos from their device storage. It is important to note that uploaded media are not compatible with *live effects*. Once the user fulfils their media input in this stage either through recording or uploading, they proceed to the next screen of the production stage that allows for further media modification. If the user did not choose any sounds in the previous stage, TikTok would attach a sound to their content **automatically**. Users can preview their work-in-progress on this page, whereas a list of options is presented on the right side of the screen, including *Edit, Templates, Text, Stickers, Effects, Filters, Add Yours, Voice*, and *Save.* By default, *Voice*, and *Save* options are hidden can could only be accessed after expanding the list. A number of features in the list stood out in our exploration: *Templates*, also referred within the app as "AutoCut", processes the video automatically based on user's selected template, and assign sounds and visual effects to the content autonomously without user inputs. Users may also browse and add a number automated visual effects from the *Effects browser* that modifies the video or specific objects within the video, such as a face, similar to *Live Effects* in the production stage. Yet unlike the previous screen, *post effects* are no longer customisable nor interactive, and offers a smaller selection that processes the video differently: their intended functionalities could be indicated by a comparison of how the two effects browsers categorise their respective effects by different tabs (**Figure 4**). While both collections offer a vast amount of options to choose from, *Live Effects* provides more recreational options, such as the "funny", "animals" and "music" categories, and offers greater navigability with search and bookmarks, while post-production effects appear to be primarily utilitarian in their functionality, and could only be browsed through manual navigation. Nevertheless, both browsers prioritise a constantly updated "trending" effects catalog, which is set as the default active selection in both screens.

---

[6] The author has attempted with up to 15 effects in a single video (which is largely uncommon in content production) and encountered no limitations in publishing. All effects appear correctly in the *effects indicator* when viewing the resulting content.

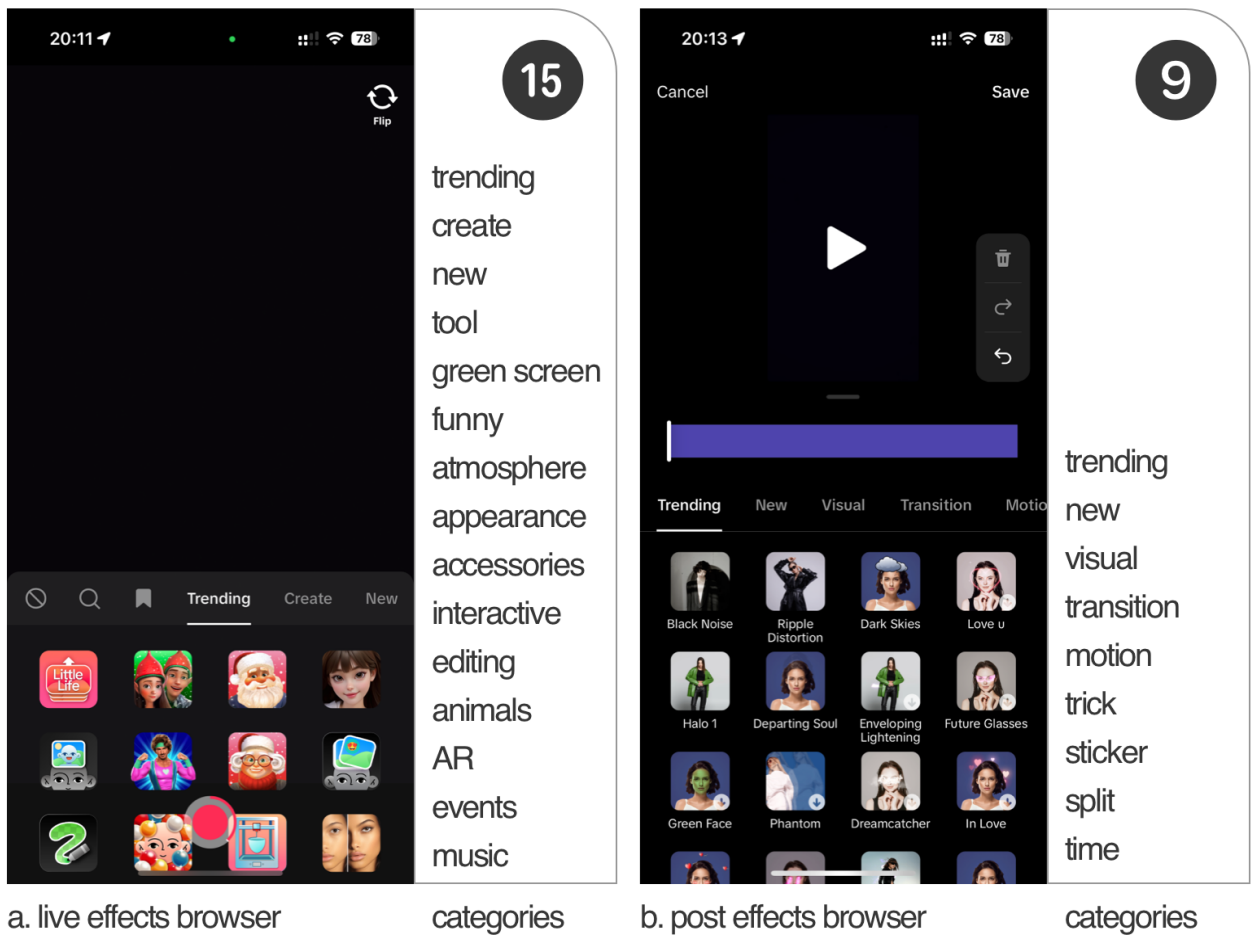

**Figure 4.** Comparison between live and post effects; the number of categories shown on top-right.

With *Stickers* and *Add Yours* options, users can introduce an interactive dimension to their content by adding elements on top of their video that respond to viewer's touches. Users can create polls, and attach sound and locations details with respective stickers. *Add Yours*, accessible either through the Sticker selector or its dedicated button, is an unique interactive sticker that invites other users to create content based on a text prompt. When users apply this sticker, they could either input their own "topic" (indicated by the sticker filler text), or choose from a list of prompts suggested by TikTok instead. By pressing the *Edit* icon, users enter a timeline editor screen for more granular adjustments to their content. The timeline editor shows the work-in-progress in a timeline layout, whereas overlaying text and effects appear below the video and audio clips as layers with adjustable duration. (Insert Screenshot of Timeline editor) Within this view, users can fine-tune their content with controls such as audio volume, video transformation, split, speed and trim, as well as automated actions, such as sound sync and virtual camera/lens movement (referred to as "Magic" in the software). Additional media may also be uploaded through the timeline editor. While the editor does not show multiple effect layers by default, several effects could be set active at the same time by users tapping on an existing effect layer and press the "add" icon. The duration of these layers are customisable.

*Distribution and Exit*

Once users finish editing their content, they can post their content to the platform either by publishing it instantly as a "Story" visible to their followers for 24 hours, or distribute the content as a TikTok video by pressing "next" and enter the publication settings page. Within publication settings, users can optionally create a description for their production and include hashtags, mentions other users or link to existing videos on TikTok. All of the effects (including both live effects and post effects) and templates applied in the content appears under the "Links" section, and is not editable for the user. Under "more options", users can choose to post their content as a template to allow other users replace the photo and video clips within the content and "reuse the same timing, transitions, sounds, effects and text" (**Figure 5**). It is also possible for other users to create stickers from the content through the "Allow Stickers" toggle, which is turned on by default if an account is public. Importantly, no matter whether the content is produced on-platform, or edited separately before uploading, the distribution setting page encourages users to attach information in the form of textual description, hashtags, location and mentioning of other users. The user completes and exits the production process when they post the content as a Story, or press the "Post" button in the publication setting page. At this point the content is uploaded to the platform and ready for circulation.

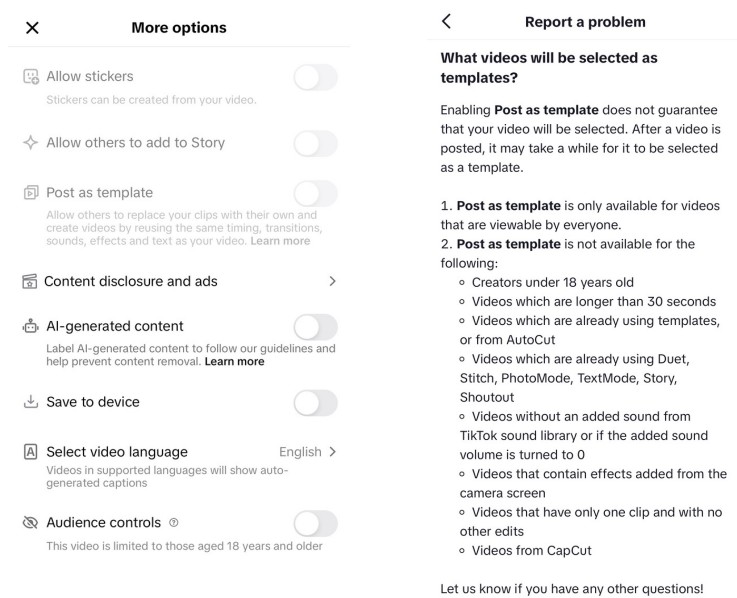

**Figure 5.** Additional Distribution options and explanations from TikTok

## DISCUSSION

Following a *data-centric, platform-aware walkthrough*, this section discusses my findings in the context of media production and platform value generation and respectively formulates two mechanisms arising from TikTok's content production: *a non-linear model of cultural production*, and platform valorisation through *a system of machinic agency*.

*The Duality of Content*

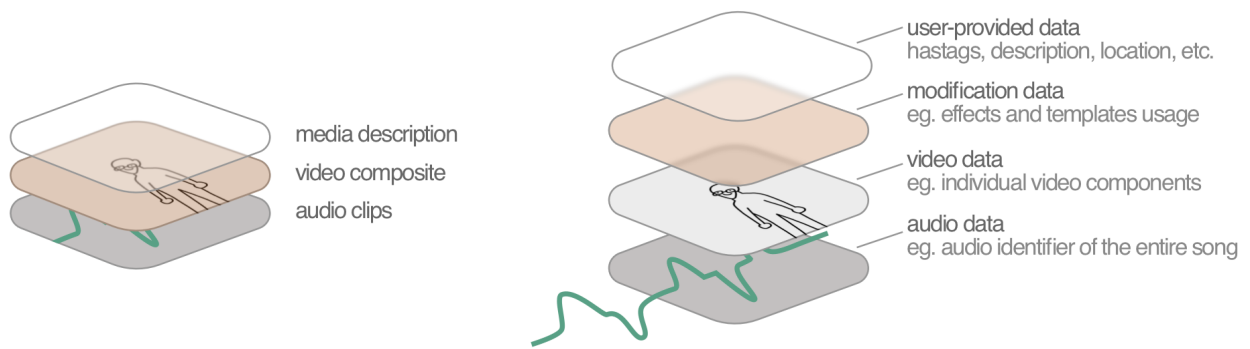

**as discursive media:**

- *flattens and objectifies* user inputs to audio-visual artefacts for consumption

- human-oriented, accessible to users

**as data assemblage:**

- *captures and preserves* information on how the content is created, what it is about, what media is used

- machine-oriented, accessible to the platform

**Figure 6.** TikTok content production yields content of a twofold nature

Starting from a broad perspective, the content production process on TikTok has been configured to create content with a twofold nature, as both *discursive media* and an *assemblage of data* that offer distinct features for the platform and its users (**Figure 6**). The former can be explicitly understood as user expressions manifest in videos that are readily available for distribution and consumption through TikTok's various viewing channels such as its algorithmic recommendation feed and direct messages. From a technological perspective, content as discursive media is best described as *flattened*[7]. Borrowing this terminology from graphic design, I am highlighting that media modifications, composites and user's expressive practices are irreversibly objectified and imbued into an audio-visual artefact in its consumable form. It is *flat* in the sense that one cannot recover the entire audio solely from a few-seconds-long fragment in a TikTok video; nor can they always identify the media modification applied to videos, as many effects can appear deceivingly natural. Regarding platform value generation, content in this notion hones potential utility to retain other users for the semantic information it contains; that is, the appeal of the content towards another user. Yet, the content's capacity to offers valuable data for TikTok's algorithm does not rely on its appeal *per se*, but rather depends on its consumption (or the user's refusal of doing so). While its semantic appeal could retain users to continue using the platform and be exposed to more advertisements, regardless its alignment with viewer preferences, the content allows TikTok to understand what the user like (or what they do not) as long as it enters users' feeds, which helps the platform to optimise its content curation for individual users.

---

[7] In graphic design softwares such as Adobe Illustrator and Procreate, *flattening* refers to the combining of layers and their components into a single one to finalise arrangements, or to reduce file size.

On the contrary, TikTok users have also been configured to create content as *data assemblage* that carries much more information beyond the visible media itself by interacting with platform technologies. While it remains challenging to understand all that TikTok captures during content production, our walkthrough has noted a few strands of data in this assemblage and their relation to platform functionalities. Apart from generic metadata that has been previously identified by scholars and developers such as textual description, location, hashtags and authorship related to the video, TikTok also captures data on how content was created and media components used in them. Counterintuitively, the presence of content data assemblages is most evident when content is circulated and consumed, during the entry stage of content production: the various UI elements (sounds, effects, templates, as well as the "add yours" prompt) overlaid on videos to provoke content-making are, in practice, an interactive representation of what content creators used to create their video. The very presence of these entry points indicates the capture and circulation of users' creative inputs on the TikTok software. When users click on entry points, they are consequently greeted with an aggregated repository of contents that also incorporate the same editing practice, whether that is sounds, effects, templates or an interactive prompt. The data assemblage can also be observed in its aggregated form through various editing features at the production stage that offer a designated "trending" category that prompts users with constantly updated popular effects and filters. These categorisations result from an algorithmic sorting and analytics of users' interaction with the content production interface, which further suggest the collection and aggregation of usage data throughout users' on-platform content generation. We could note that data collection and extraction from the user has been implemented beyond the conventional approach of capturing content consumption patterns to the capturing of production practices.

*Non-Linear Cultural Production*

My first contention underscores a non-linear model of cultural production that has been configured on TikTok in relation to the editing tools it provide. Comparing to other media platforms that primarily circulates content as discursive media, the non-linearity of cultural production on TikTok is particularly characterised by the implementation of datafication, shared resources and data-driven algorithmic sorting of these resources which now also influence the process of content production in addition to consumption.

As previously discussed, a key part of the content data assemblage is the information on *how* the content is created. The datafication of production practices is, in part, enacted by the TikTok offering of its creative options, whereas we can identify that much of the media modification and editing possibilities have been arranged so that thematically and stylistically categorised options, such as effects, music, and templates are prioritised comparing to options affording more granular modifications but does not entail specific meanings. As there lies a limited amount of options to choose from, TikTok can categorise and quantify the types of modifications users choose to incorporate in their content, thereby capturing user activities during the process of production. Editing options such as effects and sounds also actively leverage the aggregation and processing of existing content data assemblages to recommend and automatically apply algorithmically determined "trending" media to users' content. Additionally, much of TikTok's offered editing options involve some level of automation, which is best highlighted as we compare them next to traditional production environments such as Apple's Final Cut Pro, Adobe Premiere and After Effects: instead of granular edits and parameters that requires much expertise and learning in exchange of flexible and virtually endless possibilities, TikTok provides options such as effects, filters and templates that sacrifice creative granularity in exchange for an easy-to-use, oftentimes one-click access to a wide range of modifications that would otherwise require much user effort and knowledge to replicate from scratch in professional production softwares; in the case of audio resources, TikTok even automatically suggests and attaches sound to users' content-in-production when no audio is applied. Such a configuration on one hand democratises and encourages content production by reducing the amount of effort and professional knowledge required to creating content and renders the process playful through many effects' embedded interactivity, and on the other hand fosters an environment that provokes user engagement through the imitation and replication of existing content and media resources, to which scholars have coined as "imitation publics", a particular instance of networked publics that is less interpersonal or affective, but develops connections through "the shared ritual of content imitation and replication" (Zulli and Zulli 2022,1882). In both cases, TikTok's configuration effectively facilitated content production and engagement that allows the platform to generate value through their upcoming consumption and consequently user retention.

The profit-driven integration of symbolic and stylistic resources within TikTok also hold unique cultural implications. In a traditional mode of content creation, the various production process contributing to the final content is usually a *linear* process that usually yields one outcome to be distributed. The cultural significance one speaks of regarding content concerns only the final form of the content and what could be derived from it. Comparing TikTok's content circulation mechanisms next to YouTube, another video-circulating platform, we can further highlight the uniqueness of TikTok's configuration. While YouTube regulate and distribute creator-generated content, it does not offer comprehensive options to produce content but only basic adjustment tools such as trimming and caption editing. Consequently, the creators are free to use whatever tools and production environments that please them. While it could be argued that the contents hosted on the platform can be influential to the production of new videos, such influences are not applied directly by the platform, nor are they explicit, as YouTube itself does not afford the circulation of production practices pr tools for replicating content and their related resources. The cultural significance of YouTube and its hosting content is only manifest in its measures of circulating content in their *produced* form, rather than *during* the process of content-making. In the case of TikTok, content have been produced with a duality of both consumable media and data assemblage that are circulated differently to affect users as both consumers and producers. The cultural contribution of content is not only perceived in its consumption as discursive media, but also explicitly applies to users' content production by leveraging users' creative practices data to actively propose and recommend media modification, which imposes influences on users activities during the stages of content-making through its various technological affordances. We therefore coin the cultural production on TikTok as *non-linear* since its content production is configured to yield more than one typology of content that is reciprocated in both the creation and consumption of cultural artefacts following the platform's integration of both practices. We demonstrate and compare this model of TikTok in relation to an existing mode of cultural production observed on other media platforms in **Figure 7**.

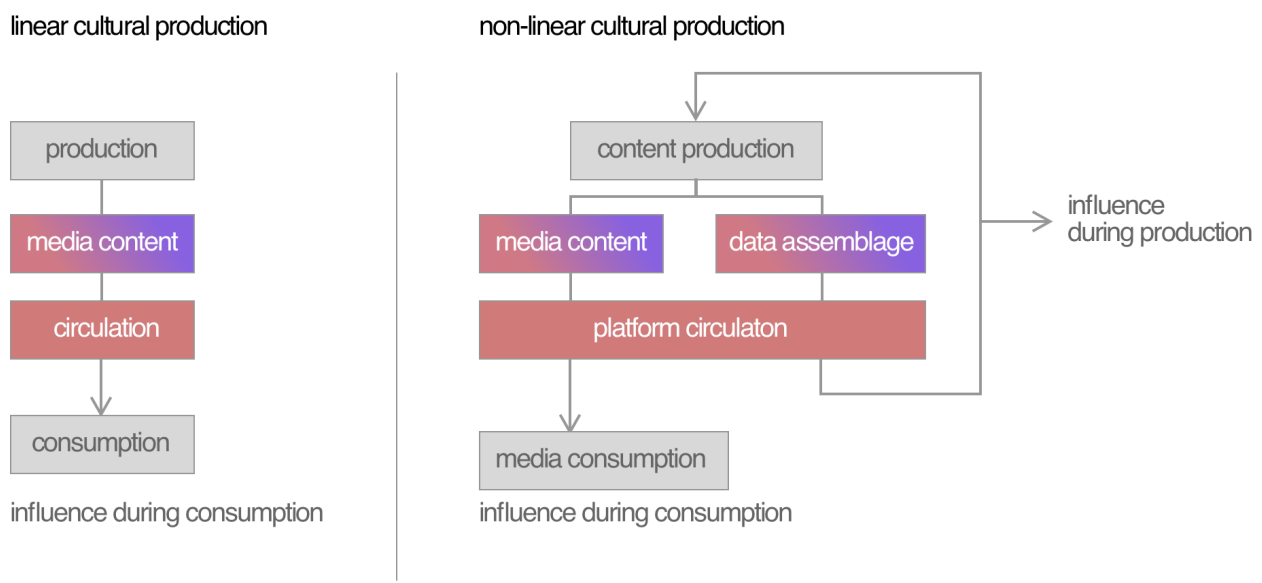

**Figure 7.** Comparison of linear and non-linear cultural production

As such, TikTok should be viewed not only as an incubator and circulator of videos, but also an actively contributor to the shaping of cultural practices that create videos to begin with. Users' media editing and modification techniques and expertise are objectified and imbued into replicable, circulating assets such as community-made effects, filters and templates. Such an arrangement give rise to a curious phenomenon, where the very *practice* of cultural production is now materialised to become an asset, next to the cultural product *resulting* from those practices.

From an economic perspective, TikTok's non-linear mode of cultural production contributes to the platform's valorisation in numerous ways. First, by simplifying content creation, it encourages users to spend more time both creating and interacting with content, which could translate into higher engagement metrics. Second, the platform's ability to gather data on user preferences during both content creation and consumption phases allows for more targeted and effective advertising strategies, thereby enhancing its revenue potential. Furthermore, this model of configuration allows TikTok content to be continuously generated in alignment with trends and popular topics by informing creators during content production. In conjunction with a similarly algorithmic-driven content distribution system, TikTok is able to sustain user interest and aggregate insights related to platform trends, which attracts advertisers and allows them to curate and optimise their promotional strategies for effective deployment and capitalisation.

*System of Machinic Agency*

Our second contention arise as we observe TikTok's valorisation scheme as a set of interactions between various parties that are both human and non-human. In this sense, I propose that the value contribution of TikTok content production is configured to emerge from its continuous interactive nature with the platform, rather than a sole matter of creativity or labour harvesting of the user in question.

As I have established in the theoretical framework, the capturing and subsequent capitalisation of users' information through their content consumption have been a dominant form of platform valorisation on TikTok. Through this lens, we have acknowledged the provision of content, amongst other factors, as an immaterial premise for platforms creating monetary value. The relation between content and platform value have brought forth the argument of affordance theorists and critical media scholars, to which they suggest the users' labour and creativity have been subject to platform exploits as a source of surplus value, whereas the exploitative nature of these activities have been concealed and disguised under the platforms' various affordances of sociality (for mechanisms of exploitation, cf. Doyle 2020; Fuchs 2013; Collie and Wilson-Barnao 2020;

on notions of affordance, cf. Ellison and Vitak 2015; boyd 2010). We find perhaps the most prominent arguments on this relation in Fuchs' work (Fuchs 2013; Fuchs and Sevignani 2013). Using a Marxist labour theory of value and extending the audience commodity theory of Dallas Smythe (1981), he poses that users have been subsumed of their labour by engaging with Internet companies, specifically their creativity in producing cultural goods as "a form of labour that is the source of the value of a data commodity that is sold to advertisers and results in profits" on platforms where users "lack the ownership and control of" (Fuchs and Sevignani 2013, 262, 288). While we can acknowledge the crucial role of users in platform economies, it is questionable whether their labour and ingenuity should be regarded as the primary source of value for two reasons. First, platforms also require and leverage users *consumption* behaviours to create monetisation opportunities, rather than the simple *creation* of content. In this sense, it has been argued that users should be more accurately situated as "productive of a *commons* that is used to extract rent from advertisers" rather than a direct, exclusive source of platform value (Rigi and Prey 2015, 403). Notions proposing user exploitation exists in multitudes beyond labour, such as their attention (Zulli 2018; Celis Bueno 2017), behaviour and experience (Zuboff 2023), and general intellect (cf. Virno and Lotringer 2007, Fumagalli et al. 2018) further exemplifies the complexity of capitalistic value generation in a contemporary setting. Second, even if we accept content production as an *ad hoc* basis of platform valorisation, as the as content circulated on social media platforms becomes more diverse and complicated in its composition, as we compare for example Facebook's textual and visual content during Fuchs time of analysis (2013) versus TikTok's focus on video content, their production now incorporates more actors and sources of agency beyond merely the creative input of users. It is in light of these concerns that I propose TikTok has configured *a system of machinic agency* to generate value for TikTok through user interactions that no longer fits definitions of work or labour, as the blurring boundary between human agency and technological contributions manifests through TikTok's content production process.

The challenge in attributing content composition and contribution to their due actors is perhaps most evident when we observe the *discursive media* products consequential to TikTok content production: it becomes incredibly difficult, if not impossible, to derive and identify whether a visual effect, audio selection or its entire conceptualisation the result of human ingenuity, or a consequence from machinic automation and algorithmic influence, as TikTok's production actively converges and obfuscate the two during every process of content production. Users could be well inspired by particular contents they encountered during their algorithm-driven consumption to enter production; effects, sounds and templates recommended during production oftentimes embeds and introduce semantic and stylistic characters to contents; in the distribution stage, users are encouraged to objectify their work in the form of templates, stories and stickers to provoke and facilitate further engagement and content-making. As such, even with an exhaustive understanding about TikTok's technical capacity, the materialist composition of contents remain largely unknown until we approach the creator themself. In fact, it is often pointless even to attribute whether that contribution is of the machines or humans, as they do not differ in the ways by which the content is consumed to retain users and generate data, while ultimately creates value for TikTok by revealing their personal interests and exposing themselves to advertisements—just to reiterate a few amongst the numerous valorisation techniques available to social media platforms. The content creation process on TikTok therefore involves a symbiotic relationship between user activities, algorithmic influence, as well as inputs from other actors such as creative professionals and sponsoring businesses to create value for the platform, rather than the sole acts of the user. After examining Marxist notions of value and its particular attribution of human agency as the only source of surplus value, Markelj and Bueno (2023) argued that in the context of datafication and increasing advancement of interactive technologies, machinic actors have also gained agency to a great extent, rending the anthropocentric approach insufficient for explaining the contribution of nonhuman agents, such as algorithms and digital technologies, in the current data economy. Informed by works of Deleuze and Guattari (1983), the authors proposed a working definition of "machines" as productive agents situated and react in a flow of information and matter. It was argued that human agency is "generated only as a result of the machinic connections that the productive unconscious forms with its environment", which is inherently related to and shaped by the organic and inorganic actors connected to them (Markelj and Bueno 2023, 13). These conceptualisations dissolve the boundary between human and non-human actors in their interactions and consequential production, which creates *machinic surplus value* that arise from their connections. By theorising agency as the ability to act and produce emerging "through dynamic encounters of machines", it usefully settles the concerns regarding the source of value and subjects of exploitation, allowing us to view TikTok as a *system of machinic agency* to account for valorisation mechanisms "that go beyond the anthropocentric definition of value grounded exclusively in human labour and human time" (ibid., 13-14). **Figure 8** visualises a system of machinic agency in the case of TikTok, with regards to different stages of content production observed in our exploration.

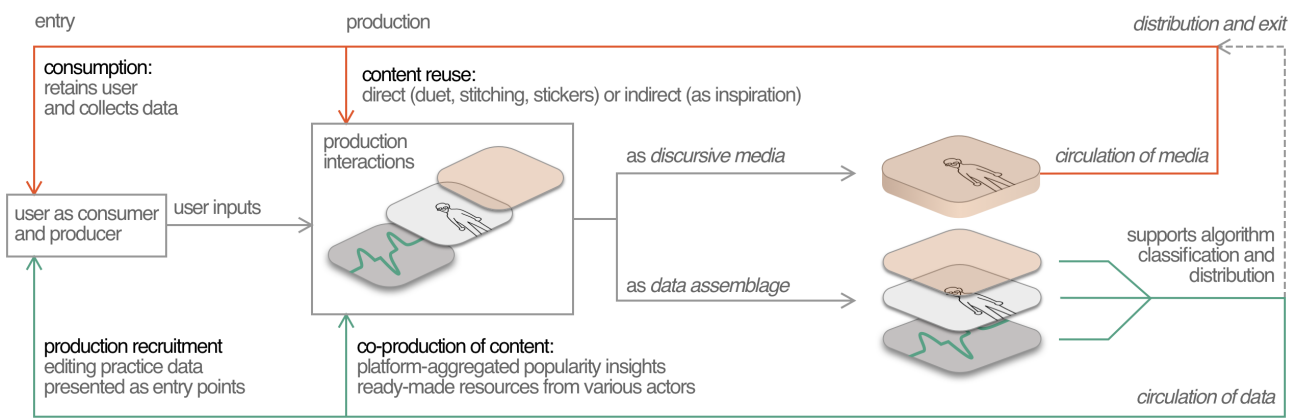

**Figure 8.** TikTok valorisation through a system of machinic agency

Through this illustration, we can identify that each stage of content production on TikTok has been configured to facilitate and affect others intricately, and contributes to the platform's valorisation with no clear distinctions of a beginning or end. For example, the consumption interface is designed to simultaneously recruit content production by repeatedly prompting users with various interactive *entry points*; users' engagement with content allows TikTok to retain users and capture their personal data; the production process incubates content as both media and data through a synthesis of user inputs, existing media as well as automation resources all aggregated within the interface, which is then circulated to facilitate value generating activities of production and consumption.

Additionally, by engaging with the user interface, content creators on TikTok also enter a classification process to convert their own production into a platform-ready dataset of semantic and stylistic indicators to optimise platform operations of content circulation. Even when users upload external media content that could have been edited already, the software nevertheless prompts users to attach trackable data (such as sounds, hashtags and location information). As a result, TikTok have designed a content production interface for the user that simultaneously functions as a video classification and data collection utility for the platform oriented towards effective algorithmic processing and enhancing capability for a personalised content delivery in user feeds. It also allows user's content and activities to be aggregated and presented in multiple ways, such as the trending sounds, effects and templates suggested during production, thereby creating a feedback loop in itself as it influences content production based on existing cultural practices of other users. This designed symbiosis of human and machinic contribution in producing content therefore extend, amplify and transform the consequences their interactions to a platform level, capable of gearing other users consumption and production practices towards profit generation. By configuring the content production process within *a system of machinic agency*, TikTok effectively operates to not only extract users of their labour, time and creativity, but also achieves valorisation through a never-ending interplay between the intellect of human users and equally provocative algorithmic technologies, deriving value from the very practice of human-computer interaction by capturing any exchanges and connections established within its integrated various sources of cultural capital and data.

# Conclusion

"Artefacts can only be understood in terms of the relation that human beings have to them." (Verbeek 2005, 117)

With its Chinese origin, TikTok has drawn as much scholarly attention as public scrutiny following its emergent popularity. Politicians and legislators have long criticised TikTok's data handling practices, political backgrounds and business ethics: ever since India banned TikTok and other Chinese apps in 2020 in defence of "sovereignty and integrity" (Schmall 2020), The United States have launched a set of federal investigations and actions regarding TikTok's Chinese affiliation and potential threat to "national interests" (cf. Clayton 2023; Maheshwari and McCabe 2024); the European Commission, EU Council and Parliament have collectively banned TikTok usage under the name of "cybersecurity concerns" (Frater 2023), with more actions in draft to address potential data abuse and user addiction (European Commission 2023; 2024). Yet at the same time, scholars and legislators have struggled to account for and assess the impacts of platforms like TikTok due to limited entry points and scarce sources of information (Lehdonvirta 2021). Like most Internet corporations, TikTok establishes a "strategic ambiguity" towards platforms operations (van der Vlist 2022, 25), concealing technologies under the name of proprietary assets and leverage narratives to obfuscate and glorify their profiting scheme. The question nowadays concerns not "whether platforms have power"; we know they do, not only from their role increasingly as the backbones of the Internet and essential "infrastructures" of our lives (cf. Helmond 2015; Schwarz 2017; Hutchinson 2023; Plantin et al. 2018), but also the unquestionable market dominance of platform-owning corporations such as Alphabet (Google, YouTube), Meta (Instagram, Facebook and WhatsApp) and Tencent (WeChat, Riot Games). The question and challenge is, as van Der Vlist (2022) promptly summarised, "where power [of digital platforms] is located precisely and how it is exercised"(133).

It is precisely these challenges that my thesis attempts to explore: with a closer look on the profiting schemes of TikTok and how it configures users' content production, we can begin to understand how the economic power of TikTok is conjured, and how its governing power is exercised to shape users practices through the software interface and contribute back towards platform valorisation.

We have begun by identifying how TikTok operates and sustains their operation in a network of stakeholders, specifically their overall reliance on advertisers as demands and users as supplies in a platform economy. From these related stakeholders, we propose that TikTok's profiting requires a configuration that facilitates maximises user retention and engagement, as well as the capturing and processing of user data, two interrelated mechanisms that mutually enhance each other to ensure the platform's optimal operation and value creation. Through the lens of platformisation, as discussed by van Dijck, Poell, and de Waal (2019), we understand that platforms intersect various economic sectors and cultural practices and leverages the user interface as a governing instrument for its valorisation-oriented user configurations. This theoretical perspective helps us conceive the multifaceted ways platforms generate value, and how practices configured by the platform could posit extensive implications across the functions and institutions it converges. In the case of TikTok, it is therefore argued that content production and consumption are closely related practices that constantly interferes with each other.

Treading through TikTok's content production interface, we are able to provide an empirical account on how platform technologies are arranged, communicated and deployed in relation to users' activities in three stages: *entry*, *production* and *distribution*. We found that TikTok not only *configures* users practices in what they can do and create, but also profoundly *transforms* the very nature of these practices with an architecture that yields content of a twofold nature, as both discursive media and data assemblages which are both processed and circulated on the platform. This dichotomy of content contributes to the platform economy through two distinct mechanisms, of which we have theorised as a *non-linear mode of cultural production* and a *system of machinic agency*. A non-linear configuration emerges from the integration of content production and consumption on TikTok, in conjunction with semantically categorised resources as data points to capture users production practices. It thus allows content to influence users at multiple stages during creation, consumption, and re-creation. Such an arrangement blurs the line between content consumption and production, and fosters cultural nexuses such as sounds, templates and effects beyond media content in their finalised form. It contributes to platform valorisation by democratising the process of content production with crowd-sourced media resources and aggregated popularity insights, which consequently drives higher engagement and allows for advertisers to engagement with content creators during the production process. Through *a system of machinic agency*, TikTok generates value through the interaction between user-machines rather than simply leveraging users' creative labour. We note from a consumption perspective that TikTok content has become a symbiotic effort of the users, communities and technologies of the platform that are oftentimes challenging, if not unfeasible to distinguish and locate these contributions. As such, the traditional Marxist theory of value, which attributes surplus value to human labor, is insufficient to explain the value generation on platforms like TikTok. Our adaptation of machinic agency suggests that both human and non-human actors (e.g., algorithms, digital technologies) contribute to the creation of value in a systematic manner. This broader perspective accounts for the dynamic interactions and productive capacities of both organic and inorganic actors within the platform's ecosystem. This configuration enables the platform to generate value not just from user labor or creativity but from the entire system of interactions facilitated by the platform and its circulation of media and data, where there lies no clear beginning or end. Such a system captures and processes data at every stage of content production and consumption to a continuous cycle of value extraction and circulation across TikTok's operations.

Following my analysis regarding these mechanisms, this thesis also invites for further studies in the following directions: As social media platforms actively contributes to users perception of reality and culture, how does the platforms' profit-driven configurations intersect and affect cultural practices? As platforms generate value through numerous approaches and leverages its networks as *systems of machinic agency*, how do we identify platforms' exploitation towards its connected actors? With the increasing application of algorithmic-driven technologies in both content creation and production, where should one locate and conceptualise creativity at the dawn of algorithmic culture? In addition to a walkthrough approach, upcoming scholars may benefit from field works towards real users, as well as systematic methodologies to unveil the algorithmic dynamics behind the visible user interface.